\documentclass[iop,apj]{emulateapj}
\usepackage{apjfonts}
\usepackage{multirow}
\usepackage{natbib}
\def\farcs{\hbox{$.\mkern-4mu^{\prime\prime}$}}

\def\hal{H$\alpha$}
\def\hb{H$\beta$}

\def\hst{{\it HST}}

\def\lambe{$\lambda_{\rm E}$}
\def\lax{{$\mathrel{\hbox{\rlap{\hbox{\lower4pt\hbox{$\sim$}}}\hbox{$<$}}}$}}
\def\gax{{$\mathrel{\hbox{\rlap{\hbox{\lower4pt\hbox{$\sim$}}}\hbox{$>$}}}$}}
\def\simlt{\lower.5ex\hbox{$\; \buildrel < \over \sim \;$}}
\def\simgt{\lower.5ex\hbox{$\; \buildrel > \over \sim \;$}}
\def\lum{erg s$^{-1}$}
\def\mbh{{$M_{\rm BH}$}}

\def\percm2{cm$^{-2}$}

\def\solmass{$M_\odot$}
\def\oii{[\ion{O}{2}]}

\def\oiii{[\ion{O}{3}]}

\def\mlb{$M_{\rm BH}-L_{\rm{bul}}$}

\def\msig{$M_{\rm BH}-\sigma_\star$}
\def\mbul{$M_{\rm bul}$}
\def\edd{$L_{{\rm bol}}$/{$L_{{\rm Edd}}$}}
\def\lbul{$L_{\rm bul}$}
\def\ser{S\'{e}rsic}

\def\rbulge{$M_{R,{\rm bul}}$}

\def\mue{$\langle \mu_e \rangle$}
\def\dmu{$\Delta \langle \mu_e \rangle$}
\def\dbul{$\Delta M_{\rm bul}$}
\shorttitle{Host Galaxies of Type 1 AGNs}
\shortauthors{KIM et al.}

\begin{document}

\title{Evidence for a Young Stellar Population in Nearby Type 1 
Active Galaxies}

\author{Minjin Kim\altaffilmark{1,2} and Luis C. Ho\altaffilmark{3,4}}

\altaffiltext{1}{Department of Astronomy and Atmospheric Sciences, 
Kyungpook National University, Daegu 702-701, Korea; mkim.astro@gmail.com}

\altaffiltext{2}{Korea Astronomy and Space Science Institute, 
Daejeon 305-348, Korea}

\altaffiltext{3}{Kavli Institute for Astronomy and Astrophysics, Peking 
University, Beijing 100871, China; lho.pku@gmail.com}

\altaffiltext{4}{Department of Astronomy, School of Physics, Peking University, Beijing 100871, China}

\begin{abstract}
To understand the physical origin of the close connection between supermassive
black holes and their host galaxies, it is vital to investigate star formation
properties in active galaxies. Using a large dataset of nearby type 1 active 
galactic nuclei (AGNs) with detailed structural decomposition based on 
high-resolution optical images obtained with the {\it Hubble Space Telescope}, 
we study the correlation between black hole mass and bulge luminosity and the 
(Kormendy) relation between bulge effective radius and surface brightness.
In both relations, the bulges of type 1 AGNs tend to be more luminous than 
those of inactive galaxies with the same black hole mass or the same bulge size.
This suggests that the central regions of AGN host galaxies have 
characteristically lower mass-to-light ratios than inactive galaxies, most 
likely due to the presence of a younger stellar population in active systems. 
In addition, the degree of luminosity excess appears to be proportional to 
the accretion rate of the AGN, revealing a physical connection between stellar 
growth and black hole growth. Adopting a simple toy model for the increase of 
stellar mass and black hole mass, we show that the fraction of young stellar 
population flattens out toward high accretion rates, possibly reflecting 
the influence of AGN-driven feedback.  
\end{abstract}

\keywords{galaxies: active --- galaxies: bulges --- galaxies: fundamental
parameters --- galaxies: photometry --- quasars: general}

\section{Introduction} 

Star formation in galaxies is often thought to be closely linked with black 
hole (BH) growth, as inferred from the observed correlation between BH mass 
and the stellar mass of the host galaxy bulge (\citealt{magorrian_1998, 
kormendy_2013}). As active galactic nuclei (AGNs) signify rapid BH growth, 
strong star formation should occur in AGN host galaxies. However, AGN activity 
and star formation occur on different timescales, and there is still 
considerable debate as to whether stellar growth and black hole growth are 
synchronized (e.g., \citealt{hickox_2014}).  

Previous observational studies have shown diverse results on the connection
between stellar growth and BH growth. Studying star formation in AGN host 
galaxies is challenging because the majority of the traditional star formation 
rate (SFR) indicators (UV, \hal, mid-infrared emission) can be heavily 
contaminated by the AGN itself.  Far-infrared (FIR) emission has been widely 
used to estimate SFR in AGN hosts, as the FIR emission from cold dust is 
thought to be dominated by star formation rather than AGNs.
Earlier FIR studies of nearby luminous AGNs 
showed that SFR appears to be tightly correlated with BH accretion rate (e.g., 
\citealt{netzer_2009b}).  However, more recent investigations,  mostly based 
on {\it Herschel}\ observations, reveal that AGNs with moderate to low 
luminosity tend to have moderate SFRs regardless of the AGN luminosity, 
revealing a weak connection between star formation and BH growth (e.g., 
\citealt{mullaney_2012, rosario_2012, rosario_2015}).  
%In addition, luminous type 2 AGNs also appear to have moderate star formation (\citealt{zakamska_2016}).
One of the main limitations of these studies is that FIR emission itself can
originate from cold dust heated by the AGN (e.g., \citealt{symeonidis_2016, 
symeonidis_2017, shangguan_2018}).  The FIR luminosity can overestimate the 
SFR in AGN hosts if the contribution from the AGN is not properly taken into 
account.   Some investigators make use of the mid-infrared emission from 
polycyclic aromatic hydrocarbons to trace star formation (e.g., 
\citealt{shi_2007, shipley_2013, alonso-herrero_2014}), but there are 
lingering doubts as to the extent to which these molecules are destroyed 
in AGN environments (e.g., \citealt{odowd_2009}). 

\citet{ho_2005} proposed that, as in normal galaxies, the luminosity of \oii\ 
$\lambda$3727 can be used to constrain the SFR in active galaxies.  Contrary to 
the results based on FIR observations, studies using \oii\ emission as a SFR 
indicator find that star formation tends to be moderately weak in nearby type 
1 (unobscured, broad-line) AGNs, regardless of their AGN luminosity  
(\citealt{kim_2006}), while star formation is enhanced in either distant type 
1 or luminous type 2 (obscured, narrow-line) AGNs (\citealt{silverman_2008, 
kim_2006}). However, dust extinction presents a potential source of uncertainty
for \oii\ emission; SFRs computed from \oii\ can be systematically 
underestimated if star formation occurs mainly in dust-enshrouded regions. 

In light of the above-mentioned technical difficulties to constrain
{\it ongoing}\ star formation in AGNs, a useful alternative strategy is 
to investigate {\it recent}\ star formation activity in AGN host galaxies. 
Analyzing the optical stellar continuum of spectra selected from the Sloan
Digital Sky Survey, \citet{kauffmann_2003} found evidence for a young 
(ages $\sim 10^8$ yr) stellar population in moderate-luminosity type 2 AGNs.  
They argued that the fraction of young stars appears to be proportional to the 
strength (luminosity) of the AGN, suggesting a close connection between stellar
growth and BH growth. On the other hand, AGN host galaxies appear to have a 
wide range of colors. Several studies show that galaxies with 
moderate-luminosity AGNs tend to have colors intermediate between those of 
red quiescent galaxies and blue star-forming galaxies (\citealt{silverman_2008,
schawinski_2009, rosario_2013a}), again revealing enhanced recent star 
formation in AGN hosts. By contrast, other studies report that the rest-frame 
optical colors of the host galaxies of moderate-luminosity AGNs are consistent 
with those of inactive galaxies (e.g., \citealt{cardamone_2010, bruce_2016}), 
as has been known to be the case for low-luminosity AGNs (\citealt{ho_2003}).
 
Notwithstanding these many previous attempts, it is still vital to better 
understand the recent star formation history of luminous type 1 AGNs, the phase 
during which BHs gain significant mass. With the advent of empirical methods 
to calculate BH masses for type 1 AGNs using single-epoch spectra (e.g., 
\citealt{kaspi_2000}), in combination with bolometric corrections to estimate
bolometric luminosities (e.g., \citealt{mcLure_2004, krawczyk_2013}), the
specific BH growth rate, defined as the mass accretion rate divided by BH mass,
can be easily inferred.  Decomposing the photometric properties of the host 
galaxies of type 1 AGNs remains a technical challenge because the bright active
nucleus often overwhelms and substantially contaminates the stellar signal.  
Thus, previous studies of the host galaxies of type 1 AGNs have been conducted 
with heterogeneous, limited samples and have reached diverse conclusions 
regarding their stellar population. Some (\citealt{sanchez_2004, canalizo_2013,
matsuoka_2014}) find that the host galaxies of luminous type 1 AGNs have bluer 
colors than normal galaxies of similar stellar mass, indicating recently 
enhanced star formation, while others (e.g., \citealt{nolan_2001, 
bettoni_2015}) disagree.

In this paper, we investigate the stellar population in nearby type 1 AGNs 
using the photometric properties of their host galaxies derived from  
{\it Hubble Space Telescope}\ (\hst) images analyzed in \citet{kim_2017}. 
We employ two independent methods---the empirical \mlb\ relation and the 
Kormendy relation---to demonstrate that AGN hosts are overluminous with 
respect to normal, inactive galaxies, an effect we attribute enhanced recent 
star formation.  The sample and data are presented in Section 2. We describe 
the scaling relations in Section 3.  We discuss the implications of the results
in Section 4 and summarize our findings in Section 5.  This work adopts the 
Vega magnitude system (\citealt{bessell_2005}) and the cosmological parameters 
$H_0 = 100 h = 67.8$ km s$^{-1}$ Mpc$^{-1}$, $\Omega_m = 0.308$, and 
$\Omega_{\Lambda} = 0.692$ (\citealt{planck_2016}).  

\section{Data}

The sample and image analysis are described in detail in \citet{kim_2017}. We 
select type 1 AGNs that have suitable optical images in the \hst\ archive 
as well as spectroscopic data either from the literature or from our own 
observations (\citealt{ho_2009}) of sufficient quality to enable their BH 
masses to be estimated (see Section 2.2).  All photometric quantities have 
been transformed to the $R$ band.  We only choose nearby objects with 
$z < 0.35$ in order to minimize evolutionary effects. Our sample includes 235 
objects, spanning a wide diversity of properties, from traditional broad-line 
Seyfert 1s and quasars to narrow-line Seyfert 1s, radio-loud and radio-quiet.
In addition to the sample from \citet{kim_2017}, we expand the dynamic range 
in physical properties by including the sample of 132 low-mass AGNs 
(\mbh\ $\leq 10^{6.3}$ \solmass) from \citet{jiang_2011}. 
Since \citet{jiang_2011} analyzed
$I$-band (F814W) images, we convert their $I$-band photometry to $R$ band 
assuming $R-I=0.65$ mag from \citet{fukugita_1995}, appropriate for Sbc 
galaxies. In total, we use a sample of 367 type 1 AGNs in this study.

\subsection{Host Properties}

We performed two-dimensional imaging analysis using {\tt GALFIT v3.0}
(\citealt{peng_2002, peng_2010}). {\tt GALFIT} allows us to decompose the 
central nucleus from the host galaxy, which is modeled with a bulge, and, if 
necessary, a disk and a bar.  The galaxy components are modeled with Fourier 
modes to accommodate complex, non-axisymmetric features such as tidal 
distortions or spiral arms.  We have also performed comprehensive and extensive
simulations to understand the measurement errors in the decomposition 
(\citealt{kim_2008a}).  We find that the luminosity ratio of the nucleus to the
underlying bulge is the main factor that determines the uncertainty of the bulge 
luminosity.  The uncertainty can increase when decomposition of bulge and disk 
is required, or when the central core of the image is saturated.  Typical 
uncertainties of the bulge luminosity range from 0.4 to 0.7 mag. $K$-correction
and color conversion from the observed filters to the $R$-band filter employ 
galaxy templates from \citet{calzetti_1994} and \citet{kinney_1996}.  
It is difficult to perform visual classification of the morphology of the host 
galaxies because the light is often overwhelmed by the bright nucleus.  
Instead, we use the measured bulge-to-total light ratio ($B/T$) and an empirical 
correlation between $B/T$ and Hubble type (\citealt{gao_2019}) to determine 
the morphological type of the hosts. 

Although our sample is relatively nearby, the bulge brightness can be affected 
by passive luminosity evolution, in the sense that more distant hosts are 
naturally more luminous because of younger stellar population.  In order to 
take into account this luminosity evolution, we adopt $dM_R/dz \approx -0.73$,
derived from a simple starburst model (\citealt{treu_2002,peng_2006a,
kim_2017}). This correction, however, barely affects the final conclusions of 
this paper.

\subsection{Black Hole Masses}

We derive BH masses using the empirical BH estimator for single-epoch 
spectra: 

\begin{eqnarray}
{\rm log} (M_{\rm BH}/M_\odot)  &=&  a + 0.533 {\rm log}
\left(\frac{L_{5100}}{10^{44} \ {\rm erg~s^{-1}}} \right) \nonumber\\ 
&+& 2 {\rm log} \left(\frac{ {\rm FWHM}} {10^3 \, {\rm km~s^{-1}}}\right),
\end{eqnarray}

\noindent
where FWHM refers to the width of the broad \hb\ line, $L_{\rm 5100}$ is the 
AGN continuum luminosity at 5100 \AA\ estimated from the nucleus magnitude 
obtained from the \hst\ image decomposition (\citealt{kim_2017}), and $a$ = 
(7.03, 6.62, 6.91) for classical bulges and ellipticals, pseudo bulges, and 
all bulge types combined, respectively (Ho \& Kim 2015).  The different 
normalizations take into account the bulge type-dependent virial coefficient 
(Ho \& Kim 2014), which reflects the fact that the \msig\ relation of inactive 
galaxies differs between classical and pseudo bulges (Kormendy \& Ho 2013).  
Whenever possible, \citet{kim_2017} classified the bulges of the AGN hosts 
using the \ser\ index and bulge-to-total light ratio [classical bulges ($n>2$ 
and $B/T > 0.2$) and pseudo bulges ($n \le 2$ or $B/T \le 0.2$)].  The bulges 
of some objects could not be classified because of the ambiguity of their 
photometric fits (e.g., merging features and dust lanes). For the sample from 
\citet{jiang_2011}, galaxies with $B/T \le 0.2$ are deemed to host pseudo 
bulges.   

The error budget for the BH mass mostly comes from the uncertainty in the 
virial factor ($\sim 0.4$ dex; \citealt{onken_2004, collin_2006, woo_2010, 
ho_2014}). Taking into account additional factors contributing to the 
uncertainty (e.g., measurement error on FWHM, intrinsic scatter of the 
broad-line region size-luminosity relation, and AGN variability), we 
conservatively adopt 0.5 dex for the uncertainty for the BH masses.  We 
estimate the bolometric luminosity and Eddington ratio (\lambe $\equiv$ \edd) 
using a bolometric correction of $L_{\rm bol} = 9.8 L_{\rm 5100}$
(\citealt{mcLure_2004}).

\begin{figure}[t]
\centering
\includegraphics[width=0.45\textwidth]{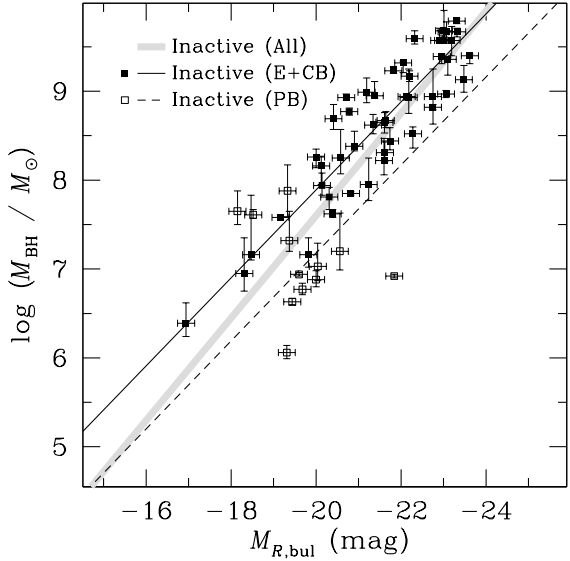}
\caption{
Correlation between BH mass and absolute $R$-band magnitude for inactive
galaxies.  Shaded grey line represents the relation for inactive galaxies,
including ellipticals, classical bulges, and pseudo bulges, adapted from
\citet{kormendy_2013}.  Filled squares and solid line denote ellipticals and
classical bulges and their relation; open circles and dashed line denote
pseudo bulges and their relation.
}
\end{figure}

\section{Scaling Relations}

\subsection{\mbh$-$\lbul\ Relation} 

To enable comparison between active and inactive galaxies, we first derive
the \mlb\ relation of inactive galaxies in the $R$ band, using the sample and 
data from \citet{kormendy_2013}. To convert bulge stellar mass to $R$-band 
luminosity, we use a mass-to-light ratio calculated from the $B-V$ colors of 
the bulges of the individual galaxies (\citealt{into_2013}). We then fit a 
linear relation of the form 

\begin{equation}
\log (M_{\rm BH}/M_\odot) = \alpha+\beta M_{R,{\rm bul}}.
\end{equation}

\noindent 
The fitting procedure adopts the $\chi^2$-minimization method of 
\citet{tremaine_2002}, which takes into account errors in both parameters.
To account for an intrinsic scatter ($\epsilon_0$) in the \mlb\ relation, 
$\chi^2$ is written as 

\begin{equation}
\chi^2 = \frac{y_i - (\alpha+\beta x_i)}
{{\epsilon_0}^2+\sigma_{y,i}+\beta \sigma_{x,i}},
\end{equation}

\noindent
where $y_i=\log (M_{\rm BH}/M_\odot)$, $x_i=M_{R,{\rm bul}}$, and 
$\sigma_{y,i}$ and $\sigma_{x,i}$ are measurement errors of $y_i$ and $x_i$, 
respectively.  The best-fit parameters for inactive galaxies are 
$\alpha=-3.92\pm0.82$, $\beta=-0.58\pm0.04$ and $\epsilon_0=0.53\pm0.06$. 

\citet{kormendy_2013} argue that the BH-host scaling relations depend on 
bulge type. To account for this effect, we derive the \mlb\ relations for two
subsamples according 
\begin{figure}[t]
\centering
\includegraphics[width=0.49\textwidth]{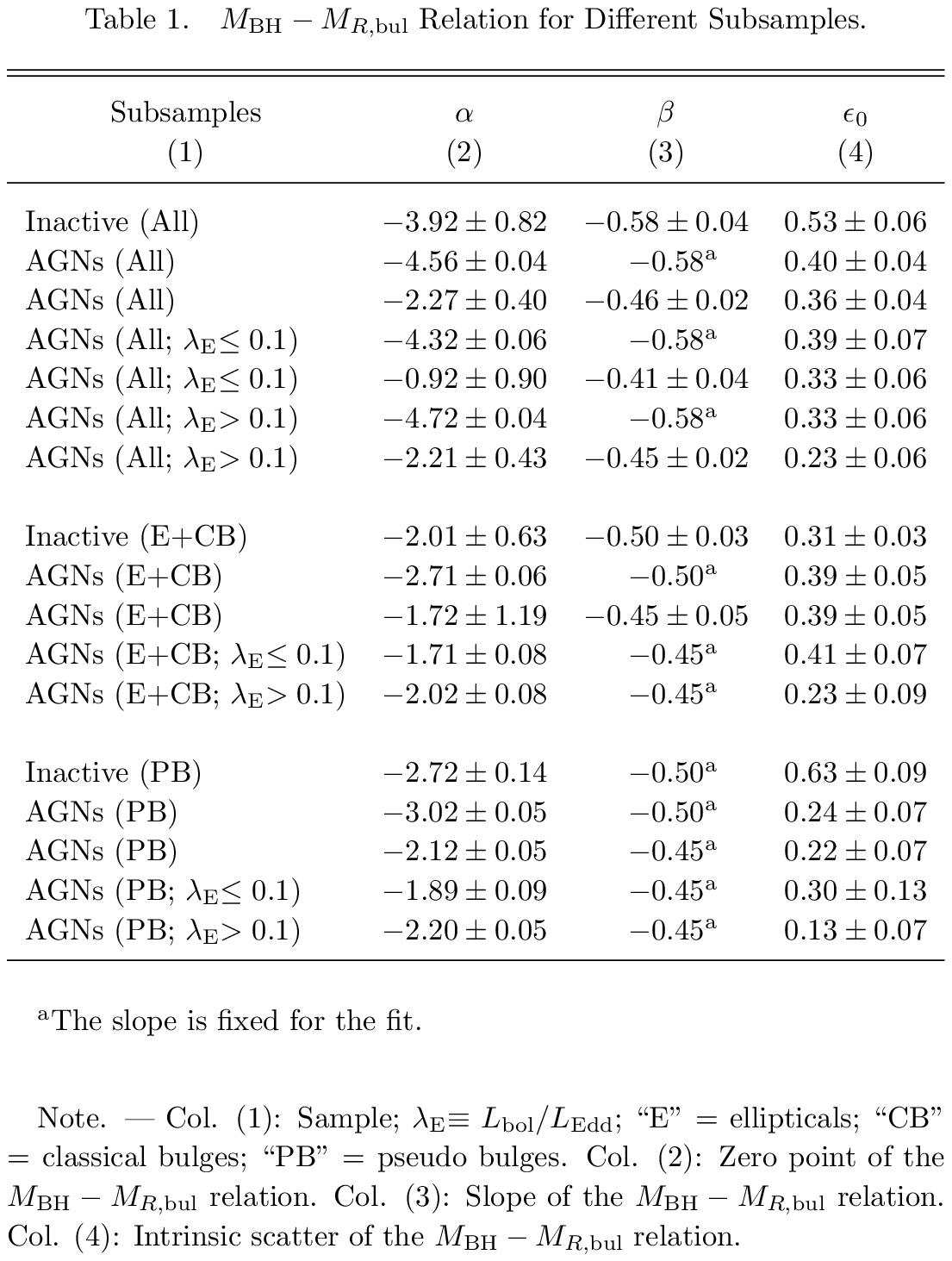}
\end{figure}
\noindent
to their bulge type, as given in \citet{kormendy_2013}.
The fit for ellipticals and classical bulges yields $\alpha=-2.01\pm0.63$, 
$\beta=-0.50\pm0.03$ and $\epsilon_0=0.31\pm0.03$.  Given that pseudo bulges 
are expected to exhibit much larger scatter (Kormendy \& Ho 2013), we do not 
fit them independently and instead fix the slope of the correlation to that of 
ellipticals and classical bulges, solving only for its zero point.  We find a 
zero point of $\alpha=-2.72\pm0.14$, indeed significantly lower than that of 
ellipticals and classical bulges (Figure 1; Table 1).

\begin{figure*}[ht!]
\includegraphics[width=0.95\textwidth]{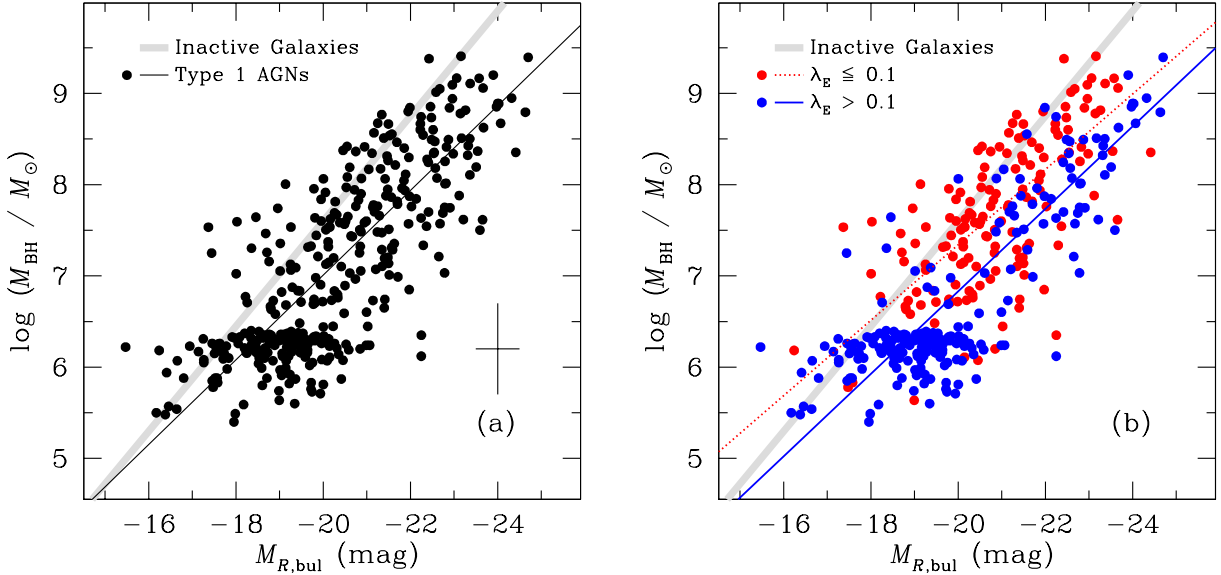}
\caption{
Correlation between BH mass and absolute $R$-band magnitude for 
the bulges of type 1 AGNs. Shaded grey line represents the relation for all
inactive galaxies regardless of their bulge type (see Fig. 1). 
(a) Solid line gives the relation for 
the entire AGN sample, with BH mass estimated using a single virial factor 
regardless of bulge types.  The typical uncertainties (0.5 dex in BH mass and 
0.5 mag in bulge magnitude) are given in the lower-right corner.  (b) The 
sample is divided according to accretion rate: red points and dotted line 
represent low Eddington ratio ($\lambda_{\rm E} \leq 0.1$); blue points and 
solid line denote high Eddington ratio ($\lambda_{\rm E} > 0.1$). 
}
\end{figure*}

\begin{figure*}[t!]
\includegraphics[width=0.95\textwidth]{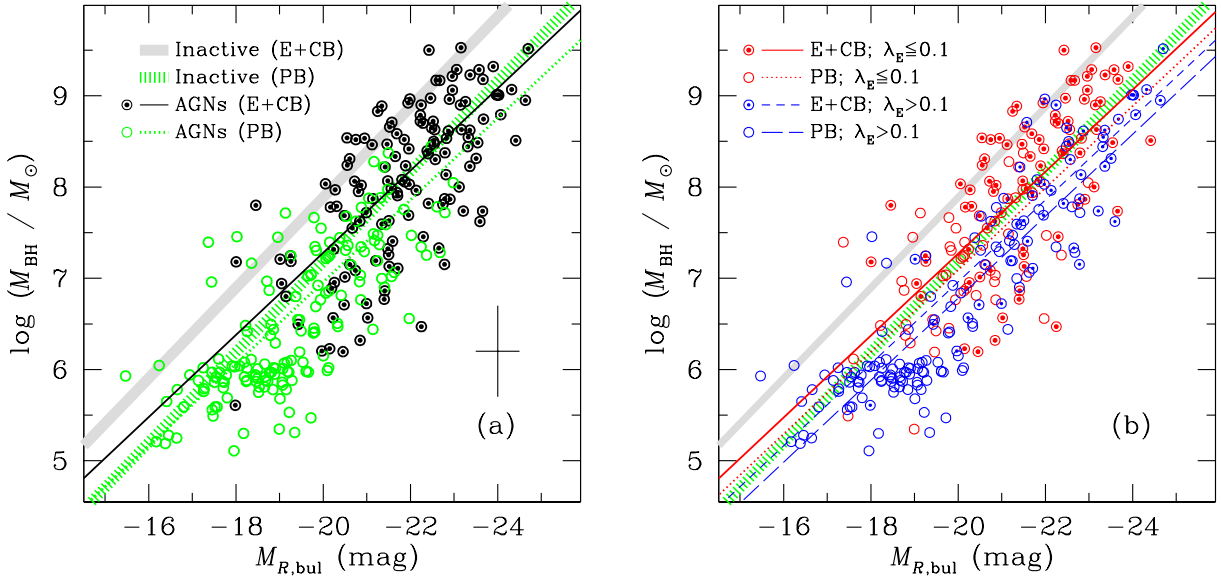}
\caption{
Same as Figure 2, except that the BH masses are estimated using different 
virial factors for pseudo bulges and classical bulges (\citealt{ho_2014}). The 
shaded grey line represents the relation for inactive ellipticals and classical
bulges from \citet{kormendy_2013}; the thick green dotted line denotes the 
relation for inactive pseudo bulges (see Section 3.1). (a) AGNs residing in 
ellipticals and classical bulges are represented by semi-filled circles and 
the solid line, while those hosted in pseudo bulges are represented by open 
circles and the dashed line.   (b) The sample is further divided according to 
accretion rate and bulge type.  Objects with low Eddington ratio 
($\lambda_{\rm E} \leq 0.1$) hosted in ellipticals and classical bulges are 
plotted as semi-filled red circles and solid line, while those in pseudo bulges
are shown as open red circles and dotted line. Objects with high Eddington 
ratio ($\lambda_{\rm E} > 0.1$) hosted in ellipticals and classical bulges are 
plotted as semi-filled blue circles and short dashed line, while those in 
pseudo bulges are shown as open blue circles and long dashed line.
}
\end{figure*}

Figures 2 and 3 show the \mlb\ relation for our type 1 AGN sample, using
different assumptions to estimate the BH mass.  As discussed in Section 2.2, 
the virial factor depends on the bulge type.  In Figure 2, we simply adopt a 
single virial factor to estimate the BH mass, assuming that AGN hosts follow 
the same \msig\ relation of inactive galaxies regardless of bulge type. The 
best fit for the AGNs yields an \mlb\ relation with $\alpha=-2.27\pm0.40$, 
$\beta=-0.46\pm0.02$ and $\epsilon_0=0.36\pm0.04$.  Although the slope of the 
\mlb\ relation of AGNs is slightly shallower than that of inactive galaxies, 
we caution against any physical interpretation of this result because our AGN 
sample is drawn from various heterogeneous \hst\ programs.  Differences in 
zero point may be more straightforward to assess.  If we fix the slope of the 
relation to that of inactive galaxies, we find that the zero point of AGNs 
is systematically lower than that of inactive galaxies, by $\sim 0.64$ dex in 
\mbh\ or $\sim 1.10$ mag in \mbul.

%\vskip 0.1in

To investigate the main driver of the offset between active and inactive 
galaxies in the \mlb\ relation, we divide the sample into two subgroups 
according to Eddington ratio ($\lambda_{\rm E}$ $\equiv$ \edd). We again find 
that the two subgroups clearly deviate in \mlb\ space, in the sense that AGNs 
with higher Eddington ratio ($\lambda_{\rm E} >  0.1$\footnote{We adopt 
this criterion because it is the median value of the sample}) tend to have 
lower BH mass or brighter bulge (Figure 2b).  

\begin{figure*}[t]
\centering
\includegraphics[width=0.95\textwidth]{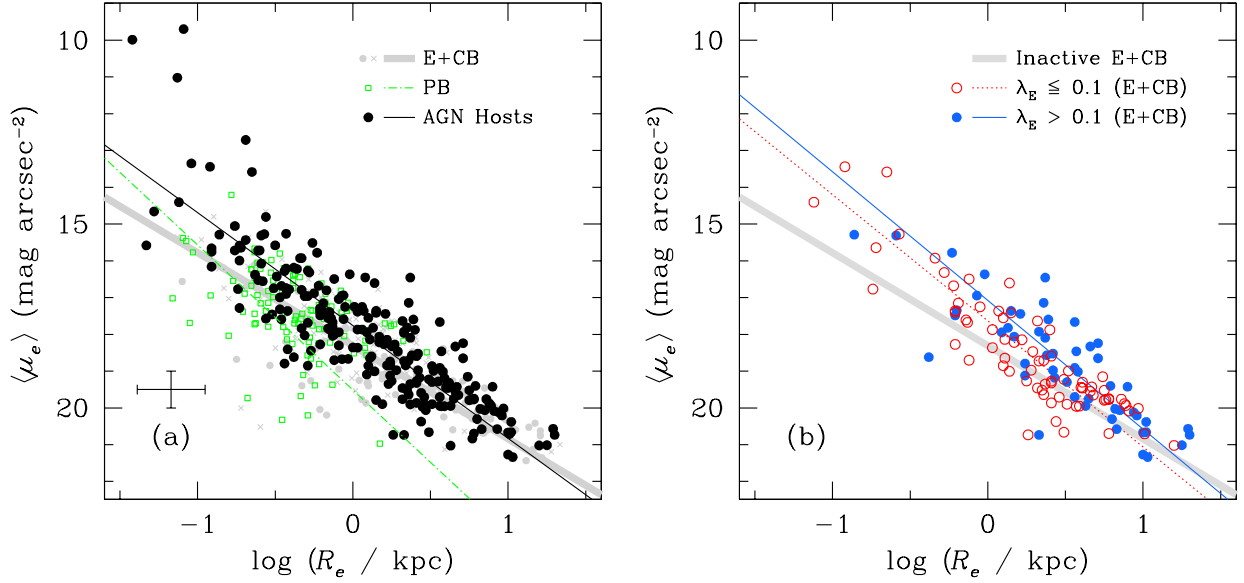}
\caption{
Correlation between effective radius $R_e$ and mean surface brightness \mue\ at 
$R_e$ (the Kormendy relation) for the bulges of AGN host galaxies.  (a) The 
filled circles and solid line represent host galaxies of AGNs irrespective of 
their bulge type.  For the comparison with inactive galaxies, we overplot the
elliptical galaxies from \citet{bender_1992}, \citet{gadotti_2009}, and 
\citet{kormendy_2009} (grey circles), disk galaxies with classical bulges from 
\citet{bender_1992}, \citet{fisher_2008}, \citet{gadotti_2009}, and 
\citet{laurikainen_2010} (grey crosses), and disk galaxies with pseudo bulges 
from \citet{fisher_2008}, \citet{gadotti_2009}, and \citet{laurikainen_2010} 
(green squares). {The relation for the inactive ellipticals and classical 
bulges is denoted by the grey line, while that for inactive pseudo bulges is 
denoted by the green dashed-dotted line. }   
The typical uncertainties (0.5 mag in \mue\ and $\sim0.2$ in $\log R_e$) 
for active galaxies are given in the lower-left corner.
(b) Here we only highlight the ellipticals and 
classical bulges, and divide the AGN hosts by Eddington ratio.  
The Kormendy relation for inactive galaxies is shown by the grey line.  
AGNs with low Eddington ratio are plotted as in red circles 
and red dotted line; those with high Eddington ratio 
are plotted in blue filled circles and blue solid line.
}
\end{figure*}

\begin{figure*}[t!]
\centering
\includegraphics[width=0.95\textwidth]{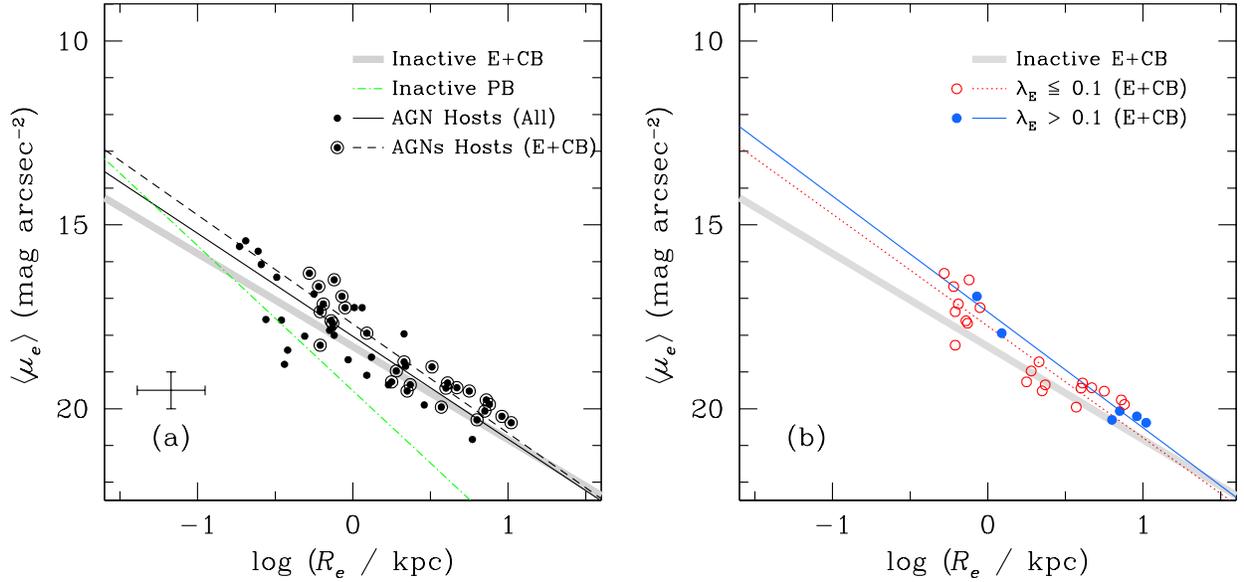}
\caption{
Same as Figure 4, except that we only plot AGN hosts with the most reliable 
measurements of bulge properties.  
}
\end{figure*}

To properly account for the fact that the \msig\ relation depends on bulge 
type, Figure 3 reexamines the correlation using BH masses calculated according 
to host galaxy bulge type.  As for the inactive galaxies, we fix the slope 
of the relation of AGNs with pseudo bulges to that of AGNs with ellipticals and 
classical bulges. At a given bulge luminosity, the BH masses of AGNs are 
significantly smaller than those of inactive galaxies; alternatively, at fixed 
BH mass AGNs have more luminous bulges than non-AGNs. For a quantitative 
comparison, we fix the slope of the relation to that of inactive ellipticals 
and classical bulges ($\beta = -0.50\pm0.03$).  The zero point of the \mlb\ 
relation for the AGN sample ($\alpha=-2.71\pm0.06$ for ellipticals and 
classical bulges; $\alpha= -3.02\pm0.05$ for pseudo bulges) is significantly 
lower than those of inactive galaxies ($\alpha=-2.01\pm0.63$ for ellipticals 
and classical bulges; $\alpha=-2.72\pm0.14$ for pseudo bulges).  It is 
interesting to note that the degree of the offset is larger for classical 
bulges ($\Delta \alpha \approx 0.7$) compared to pseudo bulges 
($\Delta \alpha \approx 0.3$).

\subsection{Kormendy Relation}

The interpretation of the observed offset between active and inactive galaxies 
in the \mlb\ relation is ambiguous.  Do AGNs have undermassive BHs or 
overluminous bulges?  Here we introduce another diagnostic that can break the 
degeneracy, one that strongly favors the possibility that active galaxies have 
overluminous bulges.  We make use of the empirical inverse correlation between 
bulge effective radius ($R_e$) and mean effective surface brightness (\mue) 
obeyed by normal galaxies.  Kormendy's (1977) relation was first introduced to 
study the bulges of early-type and S0 galaxies. Figure 4 shows the 
Kormendy relation for our AGN sample and for a sample of inactive galaxies 
assembled from the literature. The relation is derived using the ordinary 
least-squares bisector, which considers $R_e$ and \mue\ as independent 
variables. For a proper comparison between active and inactive galaxies, 
\begin{figure}[htp]
\centering
\includegraphics[width=0.48\textwidth]{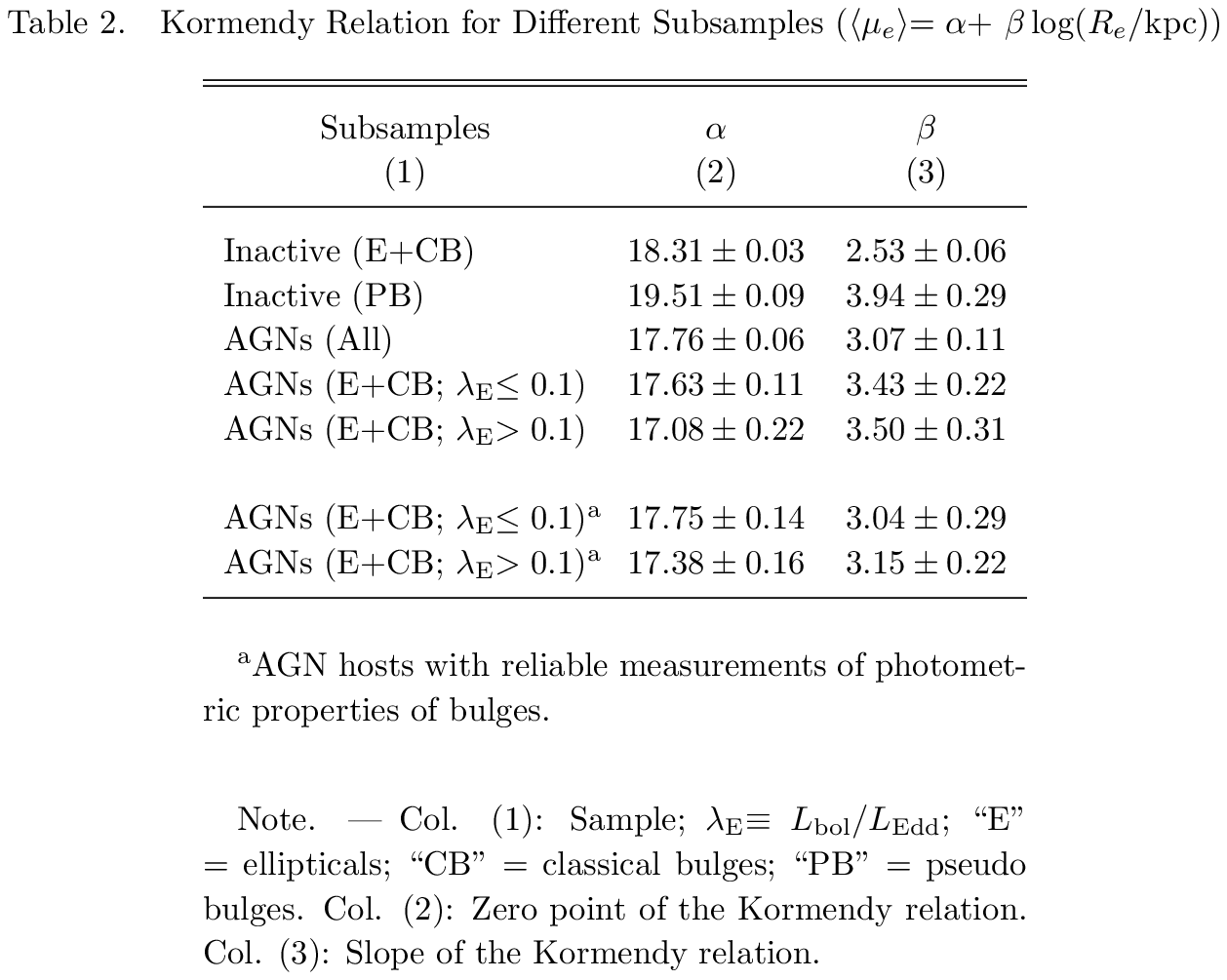}
\end{figure}
\noindent
measurement errors in \mue\ and $R_e$ are not taken into account during the 
fit, as those values are uncertain for inactive galaxies. It appears, at 
first sight, that active galaxies follow a tight relation systematically 
offset toward higher surface brightness relative to inactive galaxies.  Three 
objects (Fairall 9, $[$HB89$]$ 1549+203, and HE 0054$-$2239) have exceptionally
compact bulges ($R_e < 0.1$ kpc).  As the apparent $R_e$ of these outliers are 
comparable to the size of the point-spread function ($\sim$0\farcs1), we deem 
their size measurements to be suspicious and exclude them from further 
consideration.  

We also overplot bulge measurements for nearby inactive galaxies drawn from a 
variety of sources (\citealt{bender_1992, fisher_2008, gadotti_2009, 
kormendy_2009, laurikainen_2010}).  To estimate $R$-band surface brightness, 
we apply optical color conversions according to  morphological types from 
\cite{fukugita_1995} (e.g., $B-R=1.57$ and $V-R=0.61$ for ellipticals;
$B-R=1.39$ and $V-R=0.53$ for S0s).  For the $K$-band data from 
\citet{laurikainen_2010}, we adopt $R-K=2.56$ for S0/Sa galaxies using the 
template spectral energy distributions of \citet{polletta_2007}.
For Sloan Digital Sky Survey galaxies (\citealt{gadotti_2009}), we adopt the 
conversions from \citet{ivezic_2007}. 
Note that the control sample of 
inactive galaxies is contaminated by low-luminosity AGNs (e.g., 
\citealt{gadotti_2009}). However, AGN bolometric luminosity of 
low-luminosity AGNs ($\sim 10^{43}$ \lum) in the control sample of 
inactive galaxies inferred from a median \oiii\ luminosity 
are significantly lower than that of our sample of type 1 AGNs 
($\sim 10^{44.5}$ \lum). As a 
sanity check, we test if those low-luminosity AGNs affect the Kormendy 
relation of inactive galaxies by performing the fits with and without 
low-luminosity AGNs, and find that the change in the Kormendy relation 
is negligible. 

It is still unclear whether the Kormendy relation depends on bulge type 
(\citealt{gadotti_2009}; but see \citealt{laurikainen_2010, 
gao_2018}). Following standard convention (e.g., Kormendy \& Kennicutt 2004; 
Fisher \& Drory 2008), we divide the sample of inactive disk galaxies into two 
classes according to their bulge \ser\ index: classical bulges ($n>2$) 
and pseudo bulges ($n \le 2$).  Classical bulges follow a trend very similar 
to that of elliptical galaxies, while pseudo bulges tend to have a fainter 
zero point, larger scatter, and a seemingly steeper slope.  As the origin of 
the difference between the two bulge types is beyond the scope of this 
study, we here only focus on the Kormendy relation of ellipticals and 
classical bulges, whose intrinsic tightness provides a more useful reference 
for comparison with the AGN sample (Figure 4b).  Whereas our AGN hosts appear 
to follow a similar Kormendy relation as inactive galaxies, active galaxies 
tend to be systematically overluminous (higher \mue) compared to inactive 
early-type galaxies (ellipticals and classical bulges).  The differences 
become most pronounced at $R_e \leq\ 3$ kpc. The systematic deviation of 
AGN host galaxies from the Kormendy relation of normal, early-type galaxies 
is reminiscent of the behavior of ultraluminous infrared galaxies and quasars 
(\citealt{rothberg_2013}), here reaffirmed with a larger sample.
We also find that the degree of luminosity offset depends on accretion rate, 
in the sense that AGNs with higher Eddington ratio tend to have systematically
higher effective surface brightness.

The complexity of the image decomposition of AGNs prompts us to examine whether
systematic measurements errors or biases may artificially induce the apparent 
differences between active and inactive galaxies.  To test this hypothesis, we 
discard objects whose fits may have been compromised by a high nucleus-to-bulge
luminosity ratio, small $R_e$, strong morphological disturbance, 
point-spread function mismatch, inadequate field-of-view of the image, and dust 
obscuration.  The Kormendy relation of the remaining subset of 
58 objects, for which the errors in $\langle\mu_e\rangle$ and $R_e$ are less 
than 0.5 mag and 0.2 dex, respectively, are plotted in Figure 5. 
Although the slope and zero point have changed from the previous result for the 
entire sample, the results are qualitatively similar: AGN host galaxies have
brighter bulges than inactive galaxies. 
% and AGNs with higher accretion rates tend to have brighter bulges than AGNs 
% with lower accretion rates.  
Table 2 
summarizes the fits of the Kormendy relation for the various subsamples 
discussed above.

\begin{figure*}[t!]
\includegraphics[width=0.95\textwidth]{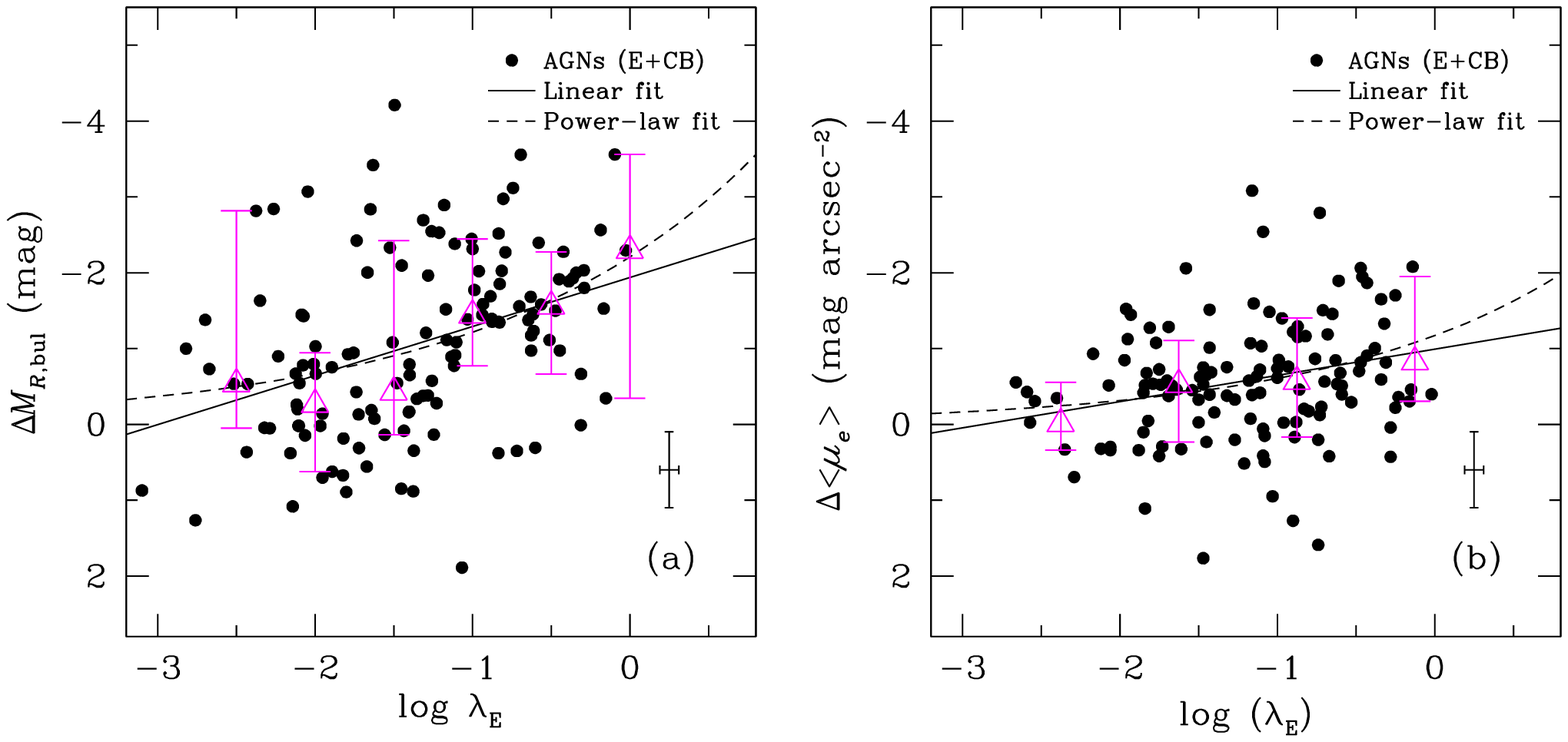}
\caption{
Dependence of excess bulge brightness on Eddington ratio ($\lambda_{\rm E}$). 
The typical uncertainties (0.5 mag in $\Delta M_{R, {\rm bul}}$ and \mue\, 
and $\sim0.06$ dex in $\lambda_{\rm E}$ are given in the lower-right corner.
(a) The amount of excess bulge luminosity ($\Delta M_{R, {\rm bul}}$) 
represents the difference between the measured \mbul\ and that inferred from 
the \mbh$-$\mbul\ relation of inactive galaxies.  (b) The amount of excess in 
mean surface brightness represents the difference between the measured \mue\ 
and that inferred from the Kormendy relation of inactive ellipticals and 
classical bulges. For both panels, the solid line indicates a linear fit, and 
the dashed line denotes a power-law fit.  The magenta triangles represent the 
median value in each bin, and error bars enclose the central 68\% of the 
values.  
}
\end{figure*}

\section{Discussion}

\subsection{The Origin of Overluminous Bulges}

Our analysis provides two lines of evidence---one based on the \mlb\ relation 
and the other on the size-surface brightness (Kormendy) relation---that the 
host galaxies of type 1 AGNs possess bulges that deviate from those of 
inactive galaxies of similar size and morphological type. Moreover, the degree 
of departure from inactive galaxies increases systematically with increasing 
Eddington ratio of the AGN.  

Taken by itself, the systematic offset between active and inactive galaxies on 
the \mlb\ relation lends itself to multiple interpretations.  One possibility 
is that active galaxies possess undermassive BHs relative to inactive galaxies 
of the same bulge luminosity.  The effect would be substantial for the current 
sample, amounting to $\sim 0.3-0.7$ dex in \mbh.  Indeed, a very similar but 
less extreme effect was noted by \citet{kim_2008b}, who, investigating a more 
limited sample of low-redshift quasar host galaxies, argued that accretion rate
is one of the primary parameters responsible for the scatter in the \mlb\ 
relation.  Ho \& Kim (2014), too, reported qualitatively similar results for a 
subset of 44 reverberation-mapped AGNs with sufficient \hst\ imaging to 
decompose the bulges of the host galaxies.  Luminous type 1 AGNs trace actively
growing BHs, and it is plausible that they might possess undermassive BHs that 
will eventually ``catch up" to the \mlb\ relation of inactive galaxies at the 
end of the AGN phase, provided, of course, that BH accretion systematically 
lags behind bulge growth.  Our sample has a median \lambe\ $\approx 0.1$.  
Assuming that these BHs grow with a constant accretion rate of 0.1\lambe\ 
during a typical AGN lifetime of $\sim$100 Myr (\citealt{martini_2004}), a 
radiative efficiency of $\epsilon=0.1$ would imply a very modest BH mass 
increase of only $\sim$0.1 dex.  The growth rate can be even smaller because 
the accretion rate of type 1 AGNs might decrease with time 
(\citealt{kelly_2010}), or the AGN lifetime may be shorter than 100 Myr 
(\citealt{martini_2004}). This falls far short of explaining the magnitude of 
the observed offset. 

Alternatively, we might posit that the BH masses suffer from some unknown 
systematic bias that caused them to be underestimated.  This seems highly 
implausible because the BH masses were estimated using a virial mass estimator 
whose virial factor was derived under the explicit assumption that 
reverberation-mapped AGNs obey the \msig\ relation of inactive galaxies.  
\citet{ho_2014} did note that the virial factor may depend on Eddington 
ratio, but the effect is mild ($\sim 0.2$ dex).  In any event, the 
reverberation-mapped AGNs used by \citet{ho_2014} to calibrate the virial 
factor have a median $\lambda_{\rm E} = 0.07$, essentially identical to that 
of the current sample (median $\lambda_{\rm E} = 0.1$).

The third and in our view most natural explanation is that AGN host galaxies 
possess {\it overluminous}\ bulges, primarily as a consequence of recent star 
formation.  For our current sample, the amount of brightening is $\sim 0.6-1.4$
mag in the $R$ band.  This interpretation is entirely consistent with previous 
studies that find that AGNs are often associated with young stellar populations 
(\citealt{kauffmann_2003, nelson_2004, letawe_2010, trump_2013, lutz_2018}). 
\citet{letawe_2010} showed that the host galaxies of nearby quasars tend to 
have colors typical of starbursts or late-type galaxies.  The host galaxies of 
the local sample of reverberation-mapped AGNs, too, appear to be brighter 
than normal galaxies at a given velocity dispersion (\citealt{nelson_2004}), 
a conclusion since confirmed by Ho \& Kim (2014) based on detailed analysis 
of their bulge component.  In principle, the bulge brightness of AGN hosts can 
also be enhanced by contamination from extended, narrow emission-line regions. 
To quantify this effect, we adopt the equivalent widths of \oiii\ $\lambda$5007
of type 2 quasars, the strongest emission line observed in these systems 
(\citealt{zakamska_2003}), which should serve as a conservative upper limit 
for type 1 AGNs.  Taking into account the FWHM of the the filter response 
functions of typical \hst\ broad-band filters ($1500-2500$ \AA), the observed 
equivalent widths of \oiii\ in type 2 quasars imply that extended narrow-line 
emission contributes less than 0.03 mag to our broad-band photometry of the 
host galaxy bulge.  This should be regarded as a strict upper limit because 
type 2 quasars on average have higher Eddington ratios ($\langle 
\lambda_{\rm E} \rangle \approx 0.2$; \citealt{kong_2018}) than our sample 
($\langle \lambda_{\rm E} \rangle \approx 0.1$). Thus, emission-line 
contamination cannot explain to the excess brightness observed in the host 
galaxy bulges.

Our analysis of the Kormendy relation of AGN host galaxies provides powerful, 
independent evidence that their bulges have enhanced brightness.  Unlike the 
\mlb\ relation, the Kormendy relation does not suffer from ambiguities 
regarding the BH mass.  The bulges of AGN host galaxies populate a Kormendy 
relation that sits systematically {\it above}\ that of normal, inactive 
galaxies.  At a given $R_e$, the bulges of AGNs have brighter \mue, and the 
effect is stronger at small $R_e$.  It is difficult to imagine that bulge size 
is the main driver for the systematic offset, because the size evolution of 
bulges is negligible for the low redshifts of our sample (e.g., 
\citealt{trujillo_2011}).  Thus, it is natural to suppose that surface 
brightness is enhanced in the bulges of AGN hosts, and the physical origin for 
the luminosity increase must be recent star formation. \citet{zhao_2019}
independently arrived at the same conclusion for a sample of low-redshift 
type 2 quasars.

For more quantitative discussion, we estimate the enhancement of the bulge 
brightness in AGN hosts compared to normal galaxies, both in the \mlb\ relation
and Kormendy relation. We define \dbul\ $\equiv$ \rbulge $-$ \rbulge(\mbh) as 
the difference between the measured bulge luminosity and the bulge luminosity 
predicted from the BH mass based on the \mlb\ relation of inactive galaxies. 
Similarly, \dmu $\equiv$ \mue $-$ \mue($R_e$) is the difference between the 
observed mean effective surface brightness and that expected from the Kormendy 
relation of inactive galaxies.   We examine the dependence of \dbul\ and \dmu\ 
on accretion rate, restricting our attention only to the subset of hosts that 
are ellipticals and classical bulges. The large intrinsic scatter of pseudo 
bulges in both scaling relations precludes any meaningful conclusions to be 
drawn for this type of bulges.  Figure 6a shows that the accretion rate is 
significantly correlated with \dbul. We estimate a Spearman correlation 
coefficient of $\rho = -0.43$ and a probability of $P_{\rm null} < 10^{-4}$ 
for the hypothesis of no correlation.  The correlation between accretion rate 
and \dmu\ is somewhat weaker, but still statistically significant ($\rho = 
-0.22$ and $P_{\rm null} \approx 0.01$).  An ordinary least squares linear
fit yields \dbul$=-0.65-1.94\log $\lambe\ and \dmu$=-0.99-0.35\log$\lambe.
It is intriguing that the fit for \dbul\ is steeper than that for \dmu, 
suggesting that the excess of bulge brightness is higher in the \mlb\ relation 
than in the Kormendy relation.  We caution, however, as discussed above, 
that systematic uncertainties in BH mass measurements may introduce
additional sources of scatter in the \mlb\ relation.  A power-law model may 
provide a better description of the data than a linear model.  We find 
\dbul$=-2.2$\lambe$^{0.26}$ and \mue$=-1.2$\lambe$^{0.29}$. The overall 
trends are similar for both fits, judging by the similarity of their root mean 
square deviation (RMSD).  For \dbul, RMSD = 21.5 for the linear fit and 21.7 
for the power-law fit; for \dmu, the corresponding values are RMSD = 9.3 and 
9.3, respectively.

\begin{figure}[t!]
\centering
\includegraphics[width=0.45\textwidth]{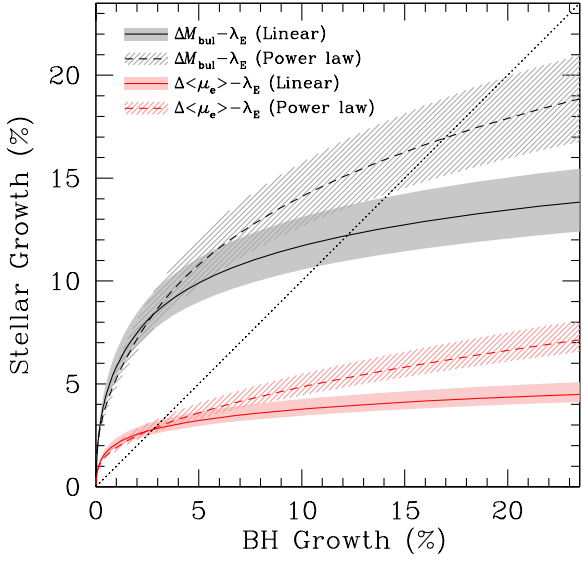}
\caption{
Comparison between BH growth rate and stellar growth rate inferred from Figure 
6. For the BH growth rate, we assume that the AGN lifetime is 50 Myr and that 
the Eddington ratio is constant over this period. For the stellar growth rate, 
we assume that the SFR is constant over a star formation lifetime of 500 Myr,
and that young stars have [Fe/H] $=0$. The relation derived from the 
\dbul$-\lambda_{\rm E}$ relation is shown as black lines, while that derived 
from the \dmu$-\lambda_{\rm E}$ relation is shown as red lines, both for the 
linear and power-law fits.   The shaded regions represent the error budget due 
to the unknown metallicity of the young stellar population. The adopted 
[Fe/H] ranges from $-2.3$ to 0.7. The dotted line represents the case where 
the BH growth rate is identical to the stellar growth rate.
}
\end{figure}
  
\subsection{Implications for the Coevolution of BHs and Their Host Galaxies}

The dependence of excess bulge brightness on Eddington ratio suggests that
star formation activity increases with increasing BH accretion rate.  This is 
naturally expected from the strong correlations between BH and bulge properties
(Kormendy \& Ho 2013), which have long been interpreted as evidence of a close 
physical connection between BH and galaxy growth, or BH accretion rate and SFR.
We employ a simple toy model to investigate how BH growth and stellar growth 
in the host galaxy are synchronized. Assuming that the excess bulge light 
arises primarily from young stars, we naturally expect that the degree of 
excess increases with increasing $f_{\rm young}$, the ratio of 
young to old stars, which is equivalent to the stellar growth
rate. In order words, the light excess is zero if the bulge contains only old 
stars (i.e. $f_{\rm young}=0$). To estimate the stellar growth rate, we assume
that the age of the old stellar population in ellipticals and classical
bulges is 10 Gyr (e.g., \citealt{zoccali_2003, brown_2006}), and that recent 
star formation occurs with a constant SFR over the duration of the star 
formation phase. The lifetime of the star formation phase is difficult to 
constrain, as it can span a wide range (100 Myr to a few Gyr) and depends on 
the origin of star formation activity (e.g., \citealt{kennicutt_1998, 
tacconi_2006, hickox_2012}).  For simplicity, we assume a star formation
lifetime of 500 Myr, but our conclusions do not depend strongly on this choice.
With these assumptions, 

\begin{equation}
{\rm stellar  \ mass\  growth\ rate} \equiv f_{\rm young} = 
\frac{M_{*,<500\, {\rm Myr}}}{M_{*, {\rm 10\,Gyr}}} \times 100,
\end{equation}

\noindent
where $M_{*,<500 \,{\rm Myr}}$ and $M_{*, {\rm 10\,Gyr}}$ are the stellar 
masses of the young and old populations, respectively. We employ a simple 
stellar population model from \cite{bruzual_2003} to calculate the 
mass-to-light ($M/L$) ratio in the $R$ band, and derive the relation between 
$M/L$ and $f_{\rm young}$.  Finally, we convert the excess bulge brightness 
(\dbul\ and \dmu) to $f_{\rm young}$.  Our fiducial model assumes that both 
old and young stars have solar metallicity, but we explore the effect of 
varying metallicity on $M/L$, denoted by the shaded regions in Figure 7.

Assuming that the Eddington ratio is constant over the lifetime of the AGN, 
the BH growth ratio can be expressed as  ${\rm exp}{\left(\lambda_{\rm E}
\frac{1-\epsilon}{\epsilon} \frac{t_{\rm AGN}}{t_{\rm Edd}}\right)}$, where 
$\epsilon = 0.1$ is the assumed radiative efficiency, $t_{\rm AGN} = 0.05$ 
Gyr is the assumed lifetime of the AGN\footnote{The actual AGN lifetime is 
somewhat uncertain (10$-$100 Myr; \citealt{martini_2004}).}, and
$t_{\rm Edd}$ is the Eddington timescale ($=0.45$ Gyr), defined as 
$M_{\rm BH}c^2/L_{\rm Edd}$ (\citealt{volonteri_2005}).  Our main 
conclusions do not depend sensitively to these adopted values.  

Figure 7 shows the results of this simple toy model, comparing the BH 
growth rate and stellar growth rate in AGN hosts as inferred from the fitted 
relations from Figure 6. As expected, the two 
quantities are clearly correlated. The absolute 
ratio between the stellar growth rate and BH growth rate has no physical 
meaning because it can depend strongly on our assumptions for the input 
parameters (e.g., lifetimes and radiative efficiency). 
Interestingly, the slope of the relation (i.e. ratio of stellar growth 
rate to BH growth rate) decreases mildly with increasing BH growth rate 
(i.e. accretion rate). 
Although it is unclear what causes this gradual change, this finding suggests 
that the {\it recent}\ specific SFR is non-linearly correlated with the 
Eddington ratio of the AGN.
It may also be consistent with the notion that star formation can be 
suppressed by AGN activity (e.g., AGN outflows; \citealt{dimatteo_2005}). 
The strength of AGN outflows is expected to be proportional to the 
accretion rate (e.g., \citealt{cicone_2014}). 

\section{Summary}

Using a new compendium of photometric parameters of the host galaxies of nearby
type 1 AGNs, we show that the bulges of the host galaxies are overluminous 
($\sim 1$ mag in the $R$ band) compared to those of inactive galaxies.  The 
effect becomes more pronounced with increasing accretion rate.  This 
is revealed both in the \mlb\ relation and in the bulge size-surface brightness 
(Kormendy) relation.  We argue that the excess bulge brightness in AGN hosts
most likely reflects their young stellar population, an interpretation 
consistent with previous, independent studies based on optical colors or 
spectra.  Our methodology, which relies solely on photometric decomposition 
and analysis of high-resolution images of the host galaxy in a single optical 
filter, demonstrates that the \mlb\ and Kormendy relations are powerful tools 
to probe the physical connection between supermassive BHs and their host 
galaxies.  We present a simple toy model to illustrate the positive correlation
between the stellar growth rate of the host galaxy and the growth rate of the 
BH.  The relative growth rate of stellar mass and BH mass decreases with 
increasing Eddington ratio, a possible manifestation of 
star formation suppression by AGN feedback.

\vskip 0.3in

We are grateful to an anonymous referee for very constructive comments.
This work was supported by the National Key Program for Science and 
Technology Research and Development (2016YFA0400702), the National Science 
Foundation of China (11721303), and the National Research Foundation 
of Korea (NRF) grant funded by the Korea government (MSIP) (No. 
2017R1C1B2002879).  We made use of the NASA/IPAC Extragalactic Database (NED), 
which is operated by the Jet Propulsion Laboratory, California Institute of 
Technology, under contract with NASA.  

\bibliography{host}
%Zhao, D., Ho, L. C., Zhao, Y., et al. 2019, \apj, submitted

\end{document}